\documentclass[acmtog,authorversion,nonacm,timestamp]{acmart}

\usepackage{caption}
\usepackage{graphicx}
\usepackage{amsmath}
\usepackage{wrapfig}

\usepackage{booktabs} 
\usepackage[ruled,vlined]{algorithm2e}
\RestyleAlgo{ruled}
\usepackage{tabularx}

\setcitestyle{square}

\def\figurePath{figures/}
\def\myfigure#1#2{\begin{figure}[ht]\centering\includegraphics*[width = \linewidth]{\figurePath#1}\caption{#2 }\label{fig:#1}\end{figure}}

\def\myWNfigure#1#2#3{\begin{wrapfigure}{r}{#2\linewidth}\centering\includegraphics*[width = \linewidth]{\figurePath#1}\label{fig:#1}\end{wrapfigure}}

\def\mySfigure#1#2#3{\begin{figure}[ht]\centering\includegraphics*[width = #2\linewidth]{\figurePath#1}\caption{#3 }\label{fig:#1}\end{figure}}

\def\mycSfigure#1#2#3{\begin{figure*}[t]\centering\includegraphics*[clip, width = #2\linewidth]{\figurePath#1}\caption{#3 }\label{fig:#1}\end{figure*}}

\newcommand{\ie}{i.e.,\ }

\newcommand{\etal}{~et~al.~}

\DeclareMathOperator*{\argmax}{arg\,max}

\usepackage{hyperref}
\usepackage{url}

\title{Closed-Loop Control of Direct Ink Writing via Reinforcement Learning}

\author{Michal Piovar\v{c}i}
\email{michael.piovarci@ist.ac.at}
\affiliation{
  \institution{ISTA}
  \country{Austria}
}
\author{Michael Foshey}
\email{mfoshey@mit.edu}
\affiliation{
  \institution{MIT CSAIL}
  \country{USA}
}
\author{Jie Xu}
\email{jiex@csail.mit.edu}
\affiliation{
  \institution{MIT CSAIL}
  \country{USA}
}
\author{Timmothy Erps}
\email{terps@csail.mit.edu}
\affiliation{
  \institution{MIT CSAIL}
  \country{USA}
}
\author{Vahid Babaei}
\email{vbabaei@mpi-inf.mpg.de}
\affiliation{
  \institution{MPI Informatics}
  \country{Germany}
}
\author{Piotr Didyk}
\email{piotr.didyk@usi.ch}
\affiliation{
  \institution{Universit\`{a} della Svizzera italiana}
  \country{Switzerland}
}
\author{Szymon Rusinkiewicz}
\email{smr@princeton.edu}
\affiliation{
  \institution{Princeton University}
  \country{USA}
}
\author{Wojciech Matusik}
\email{wojciech@csail.mit.edu}
\affiliation{
  \institution{MIT CSAIL}
  \country{USA}
}
\author{Bernd Bickel}
\email{bernd.bickel@ist.ac.at}
\affiliation{
  \institution{ISTA}
  \country{Austria}
}

\renewcommand{\shortauthors}{Piovar\v{c}i, et al.}

\keywords{closed-loop control, reinforcement learning, additive manufacturing}

\setcopyright{acmcopyright}\acmJournal{TOG}
\acmYear{2022}\acmVolume{41}\acmNumber{4}\acmArticle{112}\acmMonth{7}\acmDOI{10.1145/3528223.3530144}

\begin{document}

\begin{abstract}
Enabling additive manufacturing to employ a wide range of novel, functional materials can be a major boost to this technology. However, making such materials printable requires painstaking trial-and-error by an expert operator, as they typically tend to exhibit peculiar rheological or hysteresis properties. Even in the case of successfully finding the process parameters, there is no guarantee of print-to-print consistency due to material differences between batches. These challenges make closed-loop feedback an attractive option where the process parameters are adjusted on-the-fly. There are several challenges for designing an efficient controller: the deposition parameters are complex and highly coupled, artifacts occur after long time horizons, simulating the deposition is computationally costly, and learning on hardware is intractable. In this work, we demonstrate the feasibility of learning a closed-loop control policy for additive manufacturing using reinforcement learning. We show that approximate, but efficient, numerical simulation is sufficient as long as it allows learning the behavioral patterns of deposition that translate to real-world experiences. In combination with reinforcement learning, our model can be used to discover control policies that outperform baseline controllers. Furthermore, the recovered policies have a minimal sim-to-real gap. We showcase this by applying our control policy in-vivo on a single-layer printer using low and high viscosity materials.
\end{abstract}

\begin{teaserfigure}
  \centering
  \includegraphics{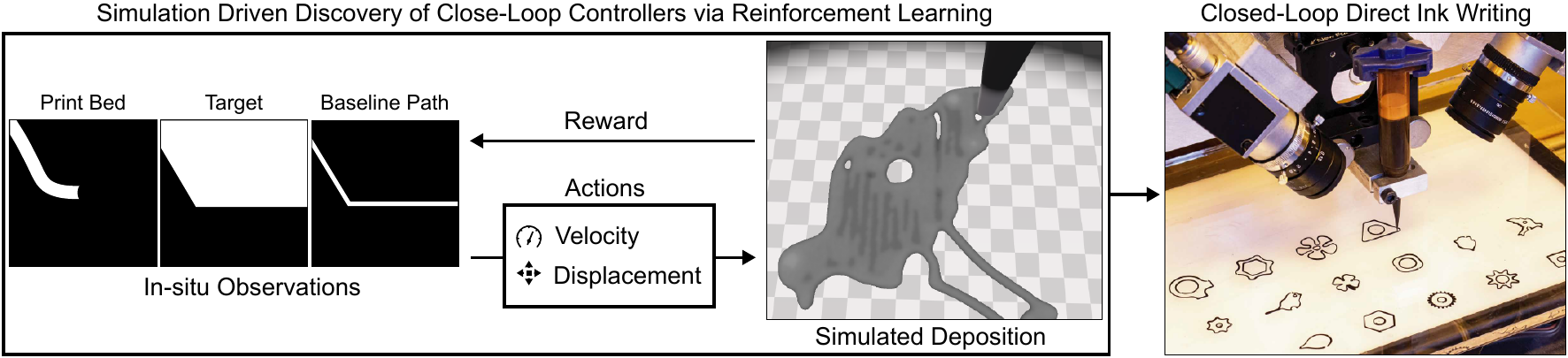}
  \caption{We propose a numerical environment suitable for learning close-loop control strategies for additive manufacturing via direct ink writing. Our method observes an in-situ view of the printing process and adjusts the velocity and printing path to achieve the desired deposition. The control policies learned exclusively in simulation can be deployed on real hardware.}
\end{teaserfigure}

\maketitle

\vspace{5ex}

\section{Introduction}

A critical component of manufacturing is identifying process parameters that consistently produce high-quality structures. In commercial devices, this is typically achieved by expensive trial-and-error experimentation~\citep{Gao2015}. To make such an optimization feasible, a critical assumption is made: the relationship between process parameters and printing outcome is predictable. However, such an assumption does not hold in practice because all manufacturing processes are stochastic in nature. Specifically, in additive manufacturing, variability in both materials and intrinsic process parameters can cause geometric errors leading to imprecision that can compromise the functional properties of the final prints. Therefore, the transition to closed-loop control is indispensable for the industrial adoption of additive manufacturing~\citep{Wang2020}.

Recently, we have seen promising progress in learning policies for interaction with amorphous materials~\citep{Li2019,Zhang2020}. Unfortunately, in the context of additive manufacturing, discovering effective control strategies is significantly more challenging. The deposition parameters have a non-linear coupling to the dynamic material properties. To assess the severity of deposition errors, we need to observe the material over long time horizons. Available simulators either lack predictive power~\citep{Mozaffar2018} or have prohibitive computational complexity for learning~\citep{Tang2018,Yan2018}. Moreover, learning on hardware is intractable as we require tens of thousands of printed samples. These challenges are further exaggerated by the limited perception of printing hardware, where typically, only a small in-situ view is available to assess the deposition quality.

In this work, we propose the first closed-loop controller for additive manufacturing trained purely in simulation that can be later deployed in real hardware. To achieve this, we formulate a custom numerical model of the deposition process. Motivated by the limited hardware perception, we make a key assumption: a numerical model is sufficiently accurate as long as we can learn behavioral patterns that hold across both simulated and real environments. This allows us to replace physically accurate but prohibitively slow simulations with efficient approximations. To ameliorate the sim-to-real gap, we enhance the simulation with a data-driven noise distribution on the spread of the deposited material. We further show that careful input and action space selection is necessary for hardware transfer. Lastly, we leverage the privileged information about the deposition process to formulate a reward function that encourages policies that account for material changes over long horizons. Thanks to the above advancements, our control policy can be trained exclusively in simulation with a minimal sim-to-real gap. We showcase this by deploying our policy on a custom single-layer direct ink writing printer. Direct ink writing is a pressure-based deposition system capable of processing a wide range of materials ranging from pastes, through hydrogels, to functional inks. Finally, we demonstrate that our policy outperforms baseline deposition methods in simulation and physical hardware with low or high viscosity materials. Code and data for this paper are at: \url{https://github.com/misop/Closed-Loop-Control-of-Direct-Ink-Writing-via-Reinforcement-Learning}.

\section{Related Work}

Many path planning strategies were devised to control the deposition process. Ranging from sparsely filling of the part interior \cite{Wu2018} to the generation of space-filling curves \cite{Zhao2016}. The key assumption of static path planners is that the printing process is reliable. Unfortunately, due to mechanical errors and material imperfections, real fabrication processes are stochastic in nature. In our work, we seek to enhance path planning with online parameter control to dynamically react to the inherent variations in the deposition.

\paragraph{Identifying Process Parameters}
Ensuring a reliable material deposition hinges on identifying stable printing parameters. To identify process parameters for additive manufacturing, it is important to understand the complex interaction between a material and a deposition process. This is typically done through trial-and-error experimentation either on hardware~\citep{Kappes2018,Wang2018,Baturynska2018b} or in simulation~\cite{Ogoke2021}. Recently, optimal experiment design and, more specifically, Gaussian processes have become a tool for efficient use of the samples to understand the deposition process~\citep{Erps2021}. However, even though Gaussian Processes model the deposition variance, they do not offer tools to adjust the deposition on-the-fly.

\paragraph{Closed-Loop Control}
Another approach to improve the printing process is to design closed-loop controllers. One of the first designs was proposed by \citet{Sitthi2015} that monitors each layer deposited by a printing process to compute an adjustment layer. \citet{Liu2017} built upon the idea and trained a discriminator that can identify the type and magnitude of observed defects. A similar approach was proposed by \citet{Yao2018} that uses handcrafted features to identify when a print significantly drops in quality. The main disadvantage of these methods is that they rely on collecting the in-situ observations to propose one corrective step by adjusting the process parameters. However, this means that the prints continue with sub-optimal parameters, and it can take several layers to adjust the deposition. In contrast, our system runs in-process and immediately reacts to the in-situ observations. This ensures high-quality deposition and adaptability to material changes.

\paragraph{Learning Closed-Loop Policies}
Recently machine learning techniques sparked a new interest in the design of adaptive control policies~\citep{Mnih2015}. A particularly successful approach for high-quality in-process control is to adopt the Model Predictive Control paradigm (MPC)~\citep{Silver2017,Oh2017,Srinivas2018,Nagabandi2018}. The control scheme of MPC relies on an observation of the current state and a short-horizon prediction of the future states. By manipulating the process parameters, we observe the changes in future predictions and can pick a future with desirable characteristics. Particularly useful is to utilize deep models to generate differentiable predictors that provide derivatives with respect to control changes~\citep{De2018,Schenck2018,Toussaint2018,Li2019a}. Unfortunately, deploying MPC on real hardware is challenging. The hardware must receive the instructions at 8 Hz for smooth control, leaving only 125 milliseconds for computation. Such a strict time budget is not sufficient for simulating the complex interactions of a deposition process. In contrast, our vision-based policies can be evaluated in under four milliseconds.

\paragraph{Policy Discovery via Reinforcement Learning}
Another option to derive control policies is to leverage deep reinforced learning~\citep{Liu2018,Peng2018,Yu2019,Lee2019,Akkaya2019}. The key challenge in the design of such controllers is formulating an efficient numerical model that captures the governing physical phenomena. As a consequence, it is most commonly applied to rigid body dynamics and rigid robots where such models are readily available~\citep{Todorov2012,Bender2014,Coumans2016,Lee2018,xu2019learning}. In contrast, learning with non-rigid objects is significantly more challenging as the computation time for deformable materials is higher and relies on some prior knowledge of the task~\citep{Clegg2018,Elliott2018,Ma2018,Wu2019}.

Closest to our work is a learning framework proposed by \citet{Zhang2020}. They demonstrate how to discover policies for human actors to manipulate visco-plastic fluids with rigid tools. While the nature of both problems is similar (interaction with materials governed by fluid dynamics), there are key differences between our methods. Zhang\etal do not consider sim-to-real transfer since their policies are not deployed in the real world. In contrast, our policies are designed to be deployed on fabrication hardware. Zhang\etal relies on full simulation state as feedback. For us, due to occlusions created by the printhead, it is infeasible to observe the full state of the deposited material. We introduce a vision system and a training method that relies on the local state to control the deposition. Their method does not learn to match a target shape. In contrast, our reward function is devised to achieve an as-close-as-possible match to the desired output while generating a uniformly flat layer. Finally, the action space proposed by Zhang\etal allows several interactions with the material. In contrast, during fabrication, materials cannot be re-adjusted after deposition. Therefore, the nature of our problem is fundamentally different as the controller needs to consider short- and long-horizon changes at deposition time.

\section{Hardware Apparatus}\label{sec:hardware}

The choice of additive manufacturing technology constraints the subsequent numerical modeling. To keep the applicability of our control policies as wide as possible, we developed a direct write 3D printing platform, (Figure~\ref{fig:hardware} left). The platform consists of a Cartesian robot with mechanical accuracy of $\pm10$ microns. We limit the acceleration to 1 mm/s$^2$ and the velocity to 2 mm/s. To smoothly control the platform, the instructions are set at 8 Hz. To deposit the material, we rely on a pressure-driven syringe pump. The syringe has a diameter of 10 microns, and the maximal pressure is 140 kPa. Such a system allows us to deposit materials from oil-like to thick pastes. To estimate the height of the deposited material, we leverage its translucency. More precisely, we correlate the deposition height with the optical intensity of our materials illuminated by a 600 lux building plate. To capture the materials, we utilize two cameras with an effective resolution of $150\times 150$ pixels. For more details about the setup, please see the appendix.

\subsection{Baseline Path Planning}\label{sec:baselinepath}

To guide the printing apparatus, we opted for an off-the-shelf slicer. The input to the slicer is a three-dimensional object. The output is a series of locations the printing head visits to reproduce the model as closely as possible. To generate a single slice of the object, we start by intersecting the 3D model with a Z-axis aligned plane (please note that this does not affect the generalizability since the input can be arbitrarily rotated). The slice is represented by a polygon that marks the outline of the printout (Figure~\ref{fig:hardware} gray). To generate the printing path, we assume a constant width of deposition (Figure~\ref{fig:hardware} red) that acts as a convolution on the printing path. The printing path (Figure~\ref{fig:hardware} blue) is created by offsetting the print boundary by half the width of the material using the Clipper algorithm~\citep{Clipper}. The infill pattern is generated by tracing a zig-zag line through the area of the print (Figure~\ref{fig:hardware} green).

\mySfigure{hardware}{0.9}{
    The printing apparatus (left) and the baseline printing policy (right).
}

\section{Reinforcement Learning for Additive Manufacturing}\label{sec:model}

The baseline path generation strictly relies on a constant width of the material. To discover policies that can adapt the printing path to the in-situ observations, we formulate the search in a reinforcement learning framework. The control problem is described by a Markov decision process $(\mathcal{S}, \mathcal{A}, \mathcal{P}, \mathcal{R})$, where $\mathcal{S}$ is the observation space, $\mathcal{A}$ is a continuous action space, $\mathcal{P} = P(s' | s, a)$ is the transition function that maps state $s$ and action $a$ to a new state $s'$, and $\mathcal{R}(s,a)\rightarrow\mathbb{R}$ is the reward function. In the following section, we will describe how to apply this framework in the context of additive manufacturing.

\subsection{Observation Space}

The choice of observation space is critical for transferring the learned knowledge from simulation to physical hardware. A natural choice would be to utilize a direct image feed from a camera module. However, the large variety of available materials would introduce significant difficulty in the learning process where materials with similar physical behavior could be treated differently based on their appearance. Moreover, the rendering would need sufficient graphical fidelity to minimize the sim-to-real gap, further limiting learning efficiency. We propose to tackle these challenges by employing an engineered observation space. Rather than using the direct appearance feed from a camera module, we process the signal into a heightmap. A heightmap is a 2D image where each pixel stores the height of the deposited material. For each height map location, the height is measured as a distance from the building plate to the deposited material. This allows our system to generalize to a wide range of sensors and materials.

We model our observation space as a small in-situ view centered at the printing nozzle. The view has a size of $3.5\times 3.5$ mm. Since the location directly under the nozzle is obscured by physical hardware, we mask a small central position equivalent to $\frac{1}{7}$th of the view. Together with the local view, we also supply the printer with the target and the baseline printing path in the local view. Increasing the data efficiency, we make the observation that the images are rotation invariant along the direction of the printer's scheduled moving path. Therefore, we align the print direction with the image X-axis. These inputs are stacked together into a 3-channel image, (Figure~\ref{fig:observation_space}). For outline printing, we threshold the heightfield to encourage a tighter fit to the target printout.

\myfigure{observation_space}{
    Input to our policy are in-situ images of the printing bed, target printout, and scheduled baseline printing path.
}

\subsection{Action Space}

\setlength{\intextsep}{0.25\intextsep}
\setlength{\columnsep}{2ex}

\myWNfigure{action_space_inset}{0.32}{
    The action space.
}
The selection of action space plays a critical role in applying the learned strategies to physical hardware. Tying the control scheme too closely to a physical setup would exaggerate the discrepancies between the physical world and our numerical model. Moreover, the learned control strategies would be valid only for particular hardware implementation and would not transfer across manufacturers of similar deposition systems. To address these issues, we propose to learn high-level control strategies. More specifically, our policy tunes the velocity of the printing head and an offset from the baseline printing path (see inset). Such a control strategy allows us to decouple high-level goals from low-level inputs. Moreover, we can lift the hardware constraints to only require an apparatus with similar capabilities. In our simulation and physical samples, we consider a velocity range of $[0.2, 2]$ mm/s and the displacement of $\pm 0.315$ mm.

\subsection{Transition Function}

In our setting, the transition function should approximate the deposition process. Unfortunately, this is a notoriously difficult problem that leads to prohibitive simulation complexities. To address this challenge, we make a key assumption: a qualitative approximation of the deposition is sufficient as long as we can learn behavior patterns that translate to real-world experiences. To achieve this goal, we propose to use an efficient numerical model. We enhance the model with a data-driven term that approximates the stochastic nature of the physical deposition. Such a combination allows us to efficiently discover control strategies that can adapt to deposition noise similar to the one observed in physical hardware.

To model the interaction of the deposited material with the printing apparatus, we rely on Position-Based Dynamics (PBD)~\cite{Muller2007}. We model the printing materials as a set of particles where each particle is defined by its position $\mathbf{p}$, velocity $\mathbf{v}$, mass $m$, and a set of constraints $C$. In our setting, we consider two constraints: (1) collision with the nozzle and (2) incompressibility of the fluid material. We model the collision with the nozzle as a hard inequality constraint:
\begin{equation}
    C_i(\mathbf{p_i}) \coloneqq (\mathbf{p_i} - \mathbf{q}_c)\cdot\mathbf{n}_c \ge 0,
\end{equation}
where $\mathbf{q}_c$ is the expected contact point of a particle with the nozzle geometry along the direction of particles motion $\mathbf{v}$ and $\mathbf{n}_c$ is the normal at the contact location. To ensure that our fluids remain incompressible, we follow \citep{Macklin2013} and formulate a density constraint for each particle:
\begin{align}
    C_i(\mathbf{p}_1,...,\mathbf{p}_n) &\coloneqq \frac{\rho_i}{\rho_0} - 1 = 0,\\
    \rho_i &= \sum_jm_jW(\mathbf{p}_i-\mathbf{p}_j,h),
\end{align}
where $\rho_0$ is the rest density, and $\rho_i$ is given by a Smoothed Particle Hydrodynamics estimator~\citep{Muller2003} in which $W$ is the smoothing kernel defined by the smoothing scale $h$.

\myWNfigure{numerical_apparatus_inset}{0.37}{
    Numerical model of the printing apparatus.
}

The physical setup contains particles in a pressurized reservoir. Unfortunately, modeling a pressurized container is computationally costly as it requires having the particles in constant contact. Instead, we approximate the deposition process at the peak of the nozzle, (inset). More specifically, we model the deposition as a particle emitter defined as a rectangle in space. The emitter generates new particles as a function of pressure:
\begin{align}
    \mathbf{x}_i &= \mathrm{distribute}(i),\;\; 0\le i\le \lfloor P\Delta t\rfloor, \\
    \mathbf{v}_i &= [0, 0, -2P]
\end{align}
where $\Delta t$ is the simulation timestep, $\mathbf{x}_i$ are newly generated particles uniformly distributed on the surface of the emitter. The number of particles and their velocity are a function of pressure $P$.

Our idealized simulation deposits material at a constant rate. However, physical hardware is not capable of such consistency. The dynamic material properties coupled with process errors introduce noise in the deposition process. To discover control strategies applicable to physical hardware, we propose reintroducing this noise into our numerical model. Due to the complex nature of the interactions between the material and the deposition process, we propose to model the printing noise in a data-driven fashion. To formulate a predictive generative model, we employ a tool from speech processing called Linear Predictive Coding (LPC)~\citep{Marple1980}. Assuming a dynamic material flow rate $\mathcal{Q}$ modeled as a time-varying function, we can predict the flow at time $N$ as a weighted sum of $M$ past flow samples and a noise term:
\begin{equation}
    \mathcal{Q}_N = -\sum_{m=1}^Ma_{M,m}\mathcal{Q}_{N-m}+\epsilon_n,
    \label{eq:noise_synth}
\end{equation}
where $\mathcal{Q}_N$ are the flow samples, $\epsilon$ is a white noise term, and $a_{M,m}$ are the parameters of $M$-th order auto-correlation filter. To find these coefficients \citet{Burg1975} propose to minimize the following energies:
\begin{align}
    e_M &= \sum_{k=1}^{N-m} |f_{M,k}|^2 + \sum_{k=1}^{N-m} |b_{M,k}|^2,\label{eq:lpc}\\
    f_{M,k} &= \sum_{i=0}^Ma_{M,i}\mathcal{Q}_{k+M-i},\\
    b_{M,k} &= \sum_{i=0}^Ma_{M,i}^*\mathcal{Q}_{k+i},
\end{align}
where $*$ denotes the complex conjugate. After finding the filter coefficients with Equation~\ref{eq:lpc} we can synthesize new width variations with similar frequency composition to the physical hardware by filtering a buzz modeled as white Gaussian noise.

To fit the model we repeatedly printed an exemplar slice (Figure~\ref{fig:LPC_model} left) and measured its width at fixed intervals (Figure~\ref{fig:LPC_model} middle). Since we sample at discrete intervals, we further smooth our data with an interpolating curve (Figure~\ref{fig:LPC_model} right). We estimated the standard deviation of the deposited material to be 175 microns, out of which 10 microns are attributed to the positional accuracy of our cartesian robot. Since the position noise is insignificant with respect to the print variations, we did not further separate the two noise terms. Finally, we experimentally estimate the magnitude of the generative noise $\epsilon$ by matching the simulation with the physical hardware.

The final predictive model generates realistic pressure variations driven by observations from the physical hardware. To incorporate these variations into our simulator, we use them to directly drive the pressure of the material emitter at time $t$ to $P_t=\mathcal{Q}_N$.

\myfigure{LPC_model}{
    We performed nine printouts and measured the width variation at specified locations. We fit the measured data with an LPC model. Please note that since our model is generative, we do not exactly match the data. Any observed resemblance is a testament to the quality of our predictor.
}

\subsection{Reward Function}

Viscous materials take significant time to settle after deposition. Therefore, it is necessary to observe the material spread over long horizons to assess deposition errors. However, the localized nature of the in-situ view does not permit such observations. At print-time, the observation window is limited by the velocity to the range of 0.875 to 8.75 seconds. In contrast, a low-viscosity material typically settles within 15 seconds of deposition. As a result, evaluating the deposition on the physical hardware is feasible only with a scan at the end of the print. In Section~\ref{sec:ablations} we will show that such a constraint on performance calculation inhibits the discovery of optimal control strategies. To learn effective control policies over long horizons, we propose to leverage the privileged information available only in the simulated environment. At each timestep, we calculate the reward as a printing performance measured on the entire printing bed. More specifically, we estimate the reward $\mathcal{R}^t$ at simulation step $t$ as:
\begin{equation}
    \mathcal{R}^t = \sum_{i,j} \mathcal{C}_{ij}\mathcal{T}_{ij} - \sum_{ij} \mathcal{C}_{ij}(1-\mathcal{T})_{ij},
\end{equation}
where $\mathcal{C}$ is the image of the printing bed and $\mathcal{T}$ is the desired target printout. The first term rewards depositing material within the target area, and the second punishes over-deposition. For infill printing, we add an additional reward term to encourage deposition with minimal height variation:
\begin{equation}
    \mathcal{R}^t = \sum_{i,j} \mathcal{C}_{ij}\mathcal{T}_{ij} - \sum_{i,j} \mathcal{C}_{ij}(1-\mathcal{T})_{ij} - std(\mathcal{C}_{ij}\mathcal{T}_{ij}).
\end{equation}
To further accelerate the learning we generate dense rewards as a delta between two steps $\mathcal{R} = \mathcal{R}^{t+1} - \mathcal{R}^t$.

\subsection{Learning Framework}

Since the proposed transition is only a rough approximation of the deposition, learning the exact model dynamics is unlikely to result in adequate control of actual hardware. Therefore, instead of learning the exact material behavior, we seek to discover behavioral patterns that enable control of material deposition. To achieve this goal, we formulate our search for control policies in a model-free framework that relies solely on in-situ observations.

Our control policy is represented as a CNN modeled after \citet{Mnih2015}. To find the policy, we follow the formulation of \citep{Schulman2017} and estimate the expected policy reward as an average over a finite batch of trials. For the training, we use a subset of the Thingy10k dataset. Each trajectory consists of fully printing a random slice. One epoch terminates by collecting 10000 observations. We run the algorithm for 4 million observations. Over time, we anneal entropy coefficient from 0.01 to 0 and the learning rate from $3\times10^{-4}$ to 0. Lastly, we picked a discount factor of 0.99, corresponding to one action having a half time of 70 steps. This is equivalent to roughly 22 mm of distance traveled. In our training set, this corresponds to 29-80 percent of the total episode length.

We experimented with training controllers for materials with varying viscosity, (Figure~\ref{fig:training_curves_noisy}). In general, we have observed that the change in viscosity did not significantly affect the learning convergence. However, we have observed a drop in performance when training control policies for deposition of low-viscosity materials. Low-viscosity materials require longer time horizons to stabilize and have a wider deposition area making precise tracing of fine features challenging. For more details about the learning process, please see the appendix.

\myfigure{training_curves_noisy}{
    Training curves for controllers with increasing viscosity.
}

\section{Results}\label{sec:results}

In this section, we provide results obtained in both virtual and physical environments. We first show that an adaptive policy can outperform baseline approaches in environments with constant deposition. Next, we showcase the in-process monitoring and the ability of our policy to adapt to dynamic environments. Finally, we demonstrate our learned controllers transferring to the physical world with a minimal sim-to-real gap.

\paragraph{Baseline Control Policy}
The baseline control uses the path planning strategy described in Section~\ref{sec:baselinepath}. To ensure consistent width between simulation and real hardware, we experimentally estimate the printing parameters by printing a series of lines with varying deposition parameters. We pick the setting that most closely matches the deposition width assumed by the path planner.

\subsection{Comparison With Baseline Controller}

We evaluate the optimized control scheme on a selection of freeform and CAD models sampled from Thingy10k~\citep{Thingi10K} and ABC~\citep{Koch2019} datasets. In total, we have 113 unseen slices corresponding to 96 unseen geometries. We report our findings in Figure~\ref{fig:evaluation_chart}. For each input slice, we report improvement on the printed boundary as the average offset. The average offset is defined as a sum of areas of under and over deposited material normalized by the outline length. More specifically, given an image of the target slice $\mathcal{T}$, printed canvas $\mathcal{C}$, and the length of the outline $l$, the average offset $\mathcal{O}$ is computed as:
\begin{equation}
    \mathcal{O} = \frac{\sum_{ij}(1 - \mathcal{C})_{ij}\mathcal{T}_{ij}}{l} + \frac{\sum_{ij}\mathcal{C}_{ij}(1-\mathcal{T})_{ij}}{l}.
\end{equation}
The improvement is calculated as a difference between the baseline and our policy. Therefore, a value higher than zero indicates that our control policy outperformed the baseline. As we can see, our policy achieved better performance in all considered models.

\mySfigure{evaluation_chart}{0.95}{
    The relative improvement of our policy over baseline in printing task with constant deposition.
}

\subsection{Performance in Dynamic Environments}\label{sec:dynenv}

We evaluate our controller in environments with stochastic material flow. To perform a quantitative evaluation, we utilize a single flow variation profile. We use the same evaluation dataset as for constant-flow policy and report the overall improvement over the baseline controller, (Figure~\ref{fig:evaluation_chart_varying}). We can observe that our closed-loop controller outperformed the baseline in each of the considered slices.

\subsubsection{Outline Printing}
To evaluate the quality of our printouts, we analyze the overflow and underflow histograms on a subset of the evaluation dataset, (Figure~\ref{fig:sim_histograms}). We can observe that the deposition histograms of our policy are narrower than the baseline. The average improvement in the standard deviation was measured at 16 microns. Moreover, the deposition histograms generated by our policy more closely approximate a normal distribution with an average skewness 0.5 lower than the baseline. We can therefore conclude that our control policy achieves a tighter control on the material deposition.

\mySfigure{evaluation_chart_varying}{0.95}{
    The relative improvement of our policy over baseline in printing task with noisy deposition.
}

\mySfigure{sim_histograms}{0.95}{
    Deposition histograms for two exemplar slices from our dataset. Even in challenging, noisy environments, our control policy achieves tighter control over the deposition process.
}

\subsubsection{Infill Printing}
We have also evaluated the infill policy in a noisy environment, (Figure~\ref{fig:infill_heightmap_inset}). We estimated the standard deviation of the deposited heightfield for the baseline at 163 microns, and our control policy at 114 microns. We can observe that the deposition noise leads to an accumulation of material. The accumulation eventually results in a bulge of material in the center of the print, complicating the deposition of subsequent layers as the material would tend to slide off. In contrast, our policy dynamically adjusts the printing path to generate a print with better height uniformity.

\mySfigure{infill_heightmap_inset}{0.95}{
    In a noisy environment, the baseline printing policy (left) significantly over-deposits and produces a bulging surface. In contrast, our policy (right) has almost no over-deposition and creates a uniform surface.
}

\subsection{Ablation Studies}\label{sec:ablations}
To demonstrate that our design decisions indeed lead to improvements in learning the deposition, we conduct several ablation studies that test the individual components: the observation space, the action space, and the reward function. Finally, we investigate how our controllers generalize across materials with different viscosities.

For each test case, we compute the average offset improvements with respect to the baseline similarly to Section~\ref{sec:dynenv}. The aggregated values are shown in Table~\ref{tab:ablations}. We treat our full controller as pre-test and ablated controllers as post-tests. We use pairwise t-tests with Holm-Bonferroni correction to estimate the statistical significance. All the evaluated components of our control policy were found to significantly improve the printing process (p-values $<0.01$).

{
\small
\begin{table}[h]
\caption{\label{tab:ablations}Average improvements of controllers from our ablation studies.}
\begin{tabular}{cccc}
\multicolumn{4}{c}{\textbf{Observation Space}}                                                                                              \\ \hline
\multicolumn{1}{c|}{Full Observation}   & \multicolumn{1}{c|}{No Print Bed} & \multicolumn{1}{c|}{No Target}      & No Path       \\
\multicolumn{1}{c|}{$\mu=9.7,\sigma=4.9$}     & \multicolumn{1}{c|}{$\mu=5.7,\sigma=7.2$}  & \multicolumn{1}{c|}{$\mu=7.2,\sigma=5.5$} & $\mu=8.4,\sigma=4.8$       \\
\end{tabular}
\begin{tabular}{ccc}
\multicolumn{3}{c}{\textbf{Action Space}}                                                                                                   \\ \hline
\multicolumn{1}{c|}{Full Action}        & \multicolumn{1}{c|}{Velocity Only}   & Displacement Only                  \\
\multicolumn{1}{c|}{$\mu=12.7,\sigma=5.7$}     & \multicolumn{1}{c|}{$\mu=7.5,\sigma=2.5$}  & $\mu=5.6,\sigma=8.3$                                            \\
\multicolumn{3}{c}{\textbf{Reward Function}}                                                                                                \\ \hline
\multicolumn{1}{c|}{Privileged Reward} & \multicolumn{1}{c|}{Delayed Reward}  & Immediate Reward                    \\
\multicolumn{1}{c|}{$\mu=12.7,\sigma=5.7$}     & \multicolumn{1}{c|}{$\mu=-22.3,\sigma=8.6$}  & $\mu=9.2,\sigma=8.0$                                           
\end{tabular}
\end{table}
}

\subsubsection{Ablation Study on Observation Space}

Our control policy relies on a live view of the deposition system to select the control parameters. However, the in-situ view is a technologically challenging addition to the printer hardware that requires a carefully calibrated setup. This ablation study verifies how vital the individual observations are to the final print quality. We consider three cases: (1) no printing bed view, (2) no target view, and (3) no future path view. Our analysis indicates that our full observation space significantly improves the quality of printouts.

\subsubsection{Ablation Study on Action Space}

\setlength{\intextsep}{0.25\intextsep}

To evaluate the need to tweak both the printing velocity and the printing path, we trained two control policies with a limited action set to either alter the velocity or the path offset. The difference in performance depends on the inherent limitations of the individual actions. On the one hand, adjusting velocity is fast (under 6.6 milliseconds) but can cope only with moderate changes in material width. This can be observed as the more prominent bulges of over-deposited material, (Figure~\ref{fig:action_ablation} dark green region). On the other hand, while offset can cope with larger material differences, it needs between 0.13 and 1.3 seconds to adjust. As a result, offset adjustment cannot cope with sudden material changes, (Figure~\ref{fig:action_ablation} light green region). In contrast, by utilizing the full action space, our policy can combine the advantages of the individual actions and minimize over-deposition.

\subsubsection{Ablation Study on Reward Function}

Our reward function uses privileged information from the numerical simulation to evaluate how the material settles over time. However, such information is not readily available on physical hardware. One either evaluates the reward once at the end of each episode to include material flow or at each timestep by disregarding long-term material motion. We evaluated how such changes to the reward function would affect our control policies. The learning process for a delayed reward is significantly slower, and it is unclear if performance similar to our policy can be achieved. On the other hand, the immediate reward policy learns faster but cannot handle material changes over longer time horizons, (Figure~\ref{fig:reward_ablation}).

\mySfigure{action_ablation}{0.9}{
   Our full action space allows adaptation to both fast (light green) and large material deviations (dark green).
}

\mySfigure{reward_ablation}{0.9}{
   Our privileged reward function facilitates the learning process resulting in improved deposition in challenging regions (green).
}

\subsubsection{Ablation Study on Viscosity}

To verify that our policy can adapt to various materials, we trained three models of varying viscosity, (Figure~\ref{fig:noisy_environment_viscorange}). We can observe that, without an adaptive control scheme, changing the material causes local over- or under-deposition. Our trained policy dynamically adjusts the offset and velocity to counterbalance the changes in the deposition. Our policy is particularly good at handling smooth width changes and quickly recovers from a spike in printing width.

We further observe how our controllers handle deviations from the training material. The policy learned for the low-viscosity materials consistently under-deposits when used to print at higher viscosities. Conversely, the control policy learned on high-viscosity material over-deposits when applied to materials with lower viscosities. From this observation, we conclude that our policy learns the spread of the material post-deposition and uses this information to guide the printing process. Therefore, slight viscosity variations are not likely to pose a significant challenge for our learned policies. However, if the learned material behavior is significantly violated, the in-situ observation space limits the ability of our policy to adapt to a before unseen material.

\subsection{Performance on Physical Hardware}

Finally, we evaluate our control policies on physical hardware. The policies were trained exclusively in simulation without any additional fine-tuning on the printing device. To conduct the evaluation, we equipped our printer with a pressure controller. The pressure control was set to a sinusoidal oscillatory signal to provide a controllable dynamic change in material properties. We used two materials with high and low viscosity, and used two separate policies pretrained in simulation using those materials. We printed 22 slices, of which 11 corresponded to the simulation training set and 11 to the evaluation set. We monitor the printing process and use the captured images to run our evaluation function to capture quantitative results. We observe that our controllers improve the average offset over the baseline print in every scenario, (Figure~\ref{fig:evaluation_chart_physical}).

\myfigure{noisy_environment_viscorange}{
    Evaluation of policy performance under varying viscosity.
}

\setlength{\intextsep}{4\intextsep}

\myfigure{evaluation_chart_physical}{
    The relative improvement of our policy over baseline in physical printing task.
}

Our control policy for high-viscosity materials achieves similar performance in both simulated (Figure~\ref{fig:evaluation_chart_varying}) and real environments (Figure~\ref{fig:evaluation_chart_physical}). We can also observe that our policy achieves a more significant improvement when handling low-viscosity materials. We attribute this improvement to the tighter control of material spread post-deposition. Lastly, we can observe that our policy for low-viscosity materials performs better on the evaluation dataset. After further investigation, there are two things we believe could cause this discrepancy. Samples with finer features are more challenging to print with low-viscosity materials. Also, the baseline performs worse on prints that require depositing a larger volume of material.

A sample of the fabricated slices can be seen in Figure~\ref{fig:physical_results_photos6_new}. The closeups show the desired deposition as green outlines. For quantitative evaluation, we overlay the prints with the target (white) and plot a histogram of under (blue) and over (red) deposited material.

We can see that our control policy transferred remarkably well to the physical hardware without any additional training. Our policy consistently achieves smaller over-deposition while not suffering from significant under-deposition. Moreover, in many cases, our policy achieves histograms with smaller widths suggesting we achieved a tighter control over the material deposition than the baseline. This demonstrates that our numerical model enables learning control policies for additive manufacturing in simulation.

\section{Limitations And Future Work}

Our control policies were trained assuming a fixed path planning strategy modeled after commercial software. However, there are other path generation strategies such as varying the infill pattern or the outline line count. In our experiments, our control policies can adjust well to different path planers at training time. However, once trained, the control policy is likely to leverage the implicitly observed path structure during execution. Therefore, for optimal results, each pathing strategy should be trained separately.

We train our controllers to deposit a single uniform layer of material. An interesting direction for future work is to extend the training to multi-layer deposition. On the one hand, multi-layer deposition brings additional freedom in correcting steps along the printing direction. On the other hand, a multi-layer printing policy needs to carefully consider material flow on the edge of each layer to minimize spilling artifacts.

Finally, we trained our controllers assuming a fixed material viscosity. In our ablation studies, we observed that a limited viscosity adaptation is possible. However, large viscosity changes require the training of separate policies. A potential line of future work is developing a system identification module to recognize the material viscosity from the in-situ view and select an appropriate controller.

\section{Conclusion}
We present a methodology for learning closed-loop control strategies for direct ink writing via reinforcement learning. To learn an effective control policy, we propose a custom numerical model of the deposition process. During the design of our model, we tackle several challenges. To obtain an efficient numerical simulator, we leverage the assumption that a numerical model is sufficiently accurate when it allows the learning of behavioral patterns that translate to the physical task. To include non-linear coupling between process parameters and printed materials, we utilize a data-driven predictive model for the deposition imperfections. Finally, to enable long horizon learning with viscous materials, we use the privileged information generated by our numerical model for reward computation. In several ablation studies, we show that these components are required to achieve high-quality printing, effectively react to instantaneous and long-horizon material changes, handle materials with varying viscosity, and adapt the deposition parameters to achieve printouts with minimal over-deposition and smooth top layers.

We demonstrate that our model can be used to train control policies that outperform baseline controllers, and transfer to physical apparatus with a minimal sim-to-real gap. We showcase this by applying control policies trained exclusively in simulation on a physical printing apparatus. We use our policies to fabricate several prototypes using low and high viscosity materials. The quantitative and qualitative analysis clearly shows the improvement of our controllers over baseline printing. This indicates that our numerical model can guide the future development of closed-loop policies for additive manufacturing. Thanks to its minimal sim-to-real gap, the model democratizes research on learning for additive manufacturing by limiting the need to invest in specialized hardware.

We optimized our control policies assuming single-layer deposition. Although this is sufficient for many applications, e.g., printed electronics, microfluidics, or bio-printing, the clear next step is multi-layer, 3D printing. To achieve this, we are improving our acquisition setup for capturing taller, more complex geometries. This will open up many new research directions, such as slicing-aware path planning and applications, such as printing optical designs, food, or functional mechanisms. We believe our introduced approach can serve as a blueprint for future research in AI-driven control of advanced manufacturing technologies, such as machining and laser-material processing. An important lesson we learned is the value of efficient numerical simulation of the involved physical phenomena. The graphics community can be a major driving force behind developing such simulations. 

\begin{acks}
This work is graciously supported by the following grant agencies: FWF Lise Meitner (Grant M 3319), SNSF (Grant 200502), ERC Starting Grant (MATERIALIZABLE-715767), NSF (Grant IIS-181507).
\end{acks}

\mycSfigure{physical_results_photos6_new}{0.94}{
    Deposition quality estimation of physical result manufactured with baseline and our learned policy.
}

\bibliographystyle{ACM-Reference-Format}
\bibliography{siggraph_conference}


\begin{thebibliography}{49}


\ifx \showCODEN    \undefined \def \showCODEN     #1{\unskip}     \fi
\ifx \showDOI      \undefined \def \showDOI       #1{#1}\fi
\ifx \showISBNx    \undefined \def \showISBNx     #1{\unskip}     \fi
\ifx \showISBNxiii \undefined \def \showISBNxiii  #1{\unskip}     \fi
\ifx \showISSN     \undefined \def \showISSN      #1{\unskip}     \fi
\ifx \showLCCN     \undefined \def \showLCCN      #1{\unskip}     \fi
\ifx \shownote     \undefined \def \shownote      #1{#1}          \fi
\ifx \showarticletitle \undefined \def \showarticletitle #1{#1}   \fi
\ifx \showURL      \undefined \def \showURL       {\relax}        \fi
\providecommand\bibfield[2]{#2}
\providecommand\bibinfo[2]{#2}
\providecommand\natexlab[1]{#1}
\providecommand\showeprint[2][]{arXiv:#2}

\bibitem[\protect\citeauthoryear{Akkaya, Andrychowicz, Chociej, Litwin, McGrew,
  Petron, Paino, Plappert, Powell, Ribas, et~al\mbox{.}}{Akkaya
  et~al\mbox{.}}{2019}]%
        {Akkaya2019}
\bibfield{author}{\bibinfo{person}{Ilge Akkaya}, \bibinfo{person}{Marcin
  Andrychowicz}, \bibinfo{person}{Maciek Chociej}, \bibinfo{person}{Mateusz
  Litwin}, \bibinfo{person}{Bob McGrew}, \bibinfo{person}{Arthur Petron},
  \bibinfo{person}{Alex Paino}, \bibinfo{person}{Matthias Plappert},
  \bibinfo{person}{Glenn Powell}, \bibinfo{person}{Raphael Ribas},
  {et~al\mbox{.}}} \bibinfo{year}{2019}\natexlab{}.
\newblock \showarticletitle{Solving rubik's cube with a robot hand}.
\newblock \bibinfo{journal}{\emph{arXiv preprint arXiv:1910.07113}}
  (\bibinfo{year}{2019}).
\newblock


\bibitem[\protect\citeauthoryear{Baturynska, Semeniuta, and
  Martinsen}{Baturynska et~al\mbox{.}}{2018}]%
        {Baturynska2018b}
\bibfield{author}{\bibinfo{person}{Ivanna Baturynska},
  \bibinfo{person}{Oleksandr Semeniuta}, {and} \bibinfo{person}{Kristian
  Martinsen}.} \bibinfo{year}{2018}\natexlab{}.
\newblock \showarticletitle{Optimization of process parameters for powder bed
  fusion additive manufacturing by combination of machine learning and finite
  element method: A conceptual framework}.
\newblock \bibinfo{journal}{\emph{Procedia Cirp}}  \bibinfo{volume}{67}
  (\bibinfo{year}{2018}), \bibinfo{pages}{227--232}.
\newblock


\bibitem[\protect\citeauthoryear{Bender, M{\"u}ller, Otaduy, Teschner, and
  Macklin}{Bender et~al\mbox{.}}{2014}]%
        {Bender2014}
\bibfield{author}{\bibinfo{person}{Jan Bender}, \bibinfo{person}{Matthias
  M{\"u}ller}, \bibinfo{person}{Miguel~A Otaduy}, \bibinfo{person}{Matthias
  Teschner}, {and} \bibinfo{person}{Miles Macklin}.}
  \bibinfo{year}{2014}\natexlab{}.
\newblock \showarticletitle{A survey on position-based simulation methods in
  computer graphics}. In \bibinfo{booktitle}{\emph{Computer graphics forum}},
  Vol.~\bibinfo{volume}{33}. Wiley Online Library, \bibinfo{pages}{228--251}.
\newblock


\bibitem[\protect\citeauthoryear{Burg}{Burg}{1975}]%
        {Burg1975}
\bibfield{author}{\bibinfo{person}{John~Parker Burg}.}
  \bibinfo{year}{1975}\natexlab{}.
\newblock \bibinfo{booktitle}{\emph{Maximum Entropy Spectral Analysis}}.
\newblock \bibinfo{publisher}{Stanford University}.
\newblock


\bibitem[\protect\citeauthoryear{Clegg, Yu, Tan, Liu, and Turk}{Clegg
  et~al\mbox{.}}{2018}]%
        {Clegg2018}
\bibfield{author}{\bibinfo{person}{Alexander Clegg}, \bibinfo{person}{Wenhao
  Yu}, \bibinfo{person}{Jie Tan}, \bibinfo{person}{C~Karen Liu}, {and}
  \bibinfo{person}{Greg Turk}.} \bibinfo{year}{2018}\natexlab{}.
\newblock \showarticletitle{Learning to dress: Synthesizing human dressing
  motion via deep reinforcement learning}.
\newblock \bibinfo{journal}{\emph{ACM Trans. Graph.}} \bibinfo{volume}{37},
  \bibinfo{number}{6} (\bibinfo{year}{2018}).
\newblock


\bibitem[\protect\citeauthoryear{Coumans and Bai}{Coumans and Bai}{2016}]%
        {Coumans2016}
\bibfield{author}{\bibinfo{person}{Erwin Coumans} {and} \bibinfo{person}{Yunfei
  Bai}.} \bibinfo{year}{2016}\natexlab{}.
\newblock \showarticletitle{Pybullet, a python module for physics simulation
  for games, robotics and machine learning}.
\newblock  (\bibinfo{year}{2016}).
\newblock


\bibitem[\protect\citeauthoryear{Elliott and Cakmak}{Elliott and
  Cakmak}{2018}]%
        {Elliott2018}
\bibfield{author}{\bibinfo{person}{Sarah Elliott} {and} \bibinfo{person}{Maya
  Cakmak}.} \bibinfo{year}{2018}\natexlab{}.
\newblock \showarticletitle{Robotic cleaning through dirt rearrangement
  planning with learned transition models}. In \bibinfo{booktitle}{\emph{ICRA
  2018}}. IEEE.
\newblock


\bibitem[\protect\citeauthoryear{Erps, Foshey, Luković, Shou, Goetzke,
  Dietsch, Stoll, von Vacano, and Matusik}{Erps et~al\mbox{.}}{2021}]%
        {Erps2021}
\bibfield{author}{\bibinfo{person}{Timothy Erps}, \bibinfo{person}{Michael
  Foshey}, \bibinfo{person}{Mina~Konaković Luković}, \bibinfo{person}{Wan
  Shou}, \bibinfo{person}{Hanns~Hagen Goetzke}, \bibinfo{person}{Herve
  Dietsch}, \bibinfo{person}{Klaus Stoll}, \bibinfo{person}{Bernhard von
  Vacano}, {and} \bibinfo{person}{Wojciech Matusik}.}
  \bibinfo{year}{2021}\natexlab{}.
\newblock \bibinfo{title}{Accelerated Discovery of 3D Printing Materials Using
  Data-Driven Multi-Objective Optimization}.
\newblock
\newblock
\showeprint[arxiv]{2106.15697}


\bibitem[\protect\citeauthoryear{Gao, Zhang, Ramanujan, Ramani, Chen, Williams,
  Wang, Shin, Zhang, and Zavattieri}{Gao et~al\mbox{.}}{2015}]%
        {Gao2015}
\bibfield{author}{\bibinfo{person}{Wei Gao}, \bibinfo{person}{Yunbo Zhang},
  \bibinfo{person}{Devarajan Ramanujan}, \bibinfo{person}{Karthik Ramani},
  \bibinfo{person}{Yong Chen}, \bibinfo{person}{Christopher~B Williams},
  \bibinfo{person}{Charlie~CL Wang}, \bibinfo{person}{Yung~C Shin},
  \bibinfo{person}{Song Zhang}, {and} \bibinfo{person}{Pablo~D Zavattieri}.}
  \bibinfo{year}{2015}\natexlab{}.
\newblock \showarticletitle{The status, challenges, and future of additive
  manufacturing in engineering}.
\newblock \bibinfo{journal}{\emph{Computer-Aided Design}}  \bibinfo{volume}{69}
  (\bibinfo{year}{2015}), \bibinfo{pages}{65--89}.
\newblock


\bibitem[\protect\citeauthoryear{Garcia, Prett, and Morari}{Garcia
  et~al\mbox{.}}{1989}]%
        {Garcia1989}
\bibfield{author}{\bibinfo{person}{Carlos~E Garcia}, \bibinfo{person}{David~M
  Prett}, {and} \bibinfo{person}{Manfred Morari}.}
  \bibinfo{year}{1989}\natexlab{}.
\newblock \showarticletitle{Model predictive control: Theory and practice—A
  survey}.
\newblock \bibinfo{journal}{\emph{Automatica}} \bibinfo{volume}{25},
  \bibinfo{number}{3} (\bibinfo{year}{1989}), \bibinfo{pages}{335--348}.
\newblock


\bibitem[\protect\citeauthoryear{Gu, Lillicrap, Sutskever, and Levine}{Gu
  et~al\mbox{.}}{2016}]%
        {Gu2016}
\bibfield{author}{\bibinfo{person}{Shixiang Gu}, \bibinfo{person}{Timothy
  Lillicrap}, \bibinfo{person}{Ilya Sutskever}, {and} \bibinfo{person}{Sergey
  Levine}.} \bibinfo{year}{2016}\natexlab{}.
\newblock \showarticletitle{Continuous deep q-learning with model-based
  acceleration}. In \bibinfo{booktitle}{\emph{International conference on
  machine learning}}. PMLR, \bibinfo{pages}{2829--2838}.
\newblock


\bibitem[\protect\citeauthoryear{Johnson}{Johnson}{2015}]%
        {Clipper}
\bibfield{author}{\bibinfo{person}{Angus Johnson}.}
  \bibinfo{year}{2015}\natexlab{}.
\newblock \bibinfo{title}{{Clipper - an open source freeware library for
  clipping and offsetting lines and polygons}}.
\newblock
  \bibinfo{howpublished}{\url{http://www.angusj.com/delphi/clipper.php}}.
\newblock


\bibitem[\protect\citeauthoryear{Kappes, Moorthy, Drake, Geerlings, and
  Stebner}{Kappes et~al\mbox{.}}{2018}]%
        {Kappes2018}
\bibfield{author}{\bibinfo{person}{Branden Kappes},
  \bibinfo{person}{Senthamilaruvi Moorthy}, \bibinfo{person}{Dana Drake},
  \bibinfo{person}{Henry Geerlings}, {and} \bibinfo{person}{Aaron Stebner}.}
  \bibinfo{year}{2018}\natexlab{}.
\newblock \showarticletitle{Machine learning to optimize additive manufacturing
  parameters for laser powder bed fusion of Inconel 718}. In
  \bibinfo{booktitle}{\emph{Proceedings of the 9th International Symposium on
  Superalloy 718 \& Derivatives: Energy, Aerospace, and Industrial
  Applications}}. Springer, \bibinfo{pages}{595--610}.
\newblock


\bibitem[\protect\citeauthoryear{Koch, Matveev, Jiang, Williams, Artemov,
  Burnaev, Alexa, Zorin, and Panozzo}{Koch et~al\mbox{.}}{2019}]%
        {Koch2019}
\bibfield{author}{\bibinfo{person}{Sebastian Koch}, \bibinfo{person}{Albert
  Matveev}, \bibinfo{person}{Zhongshi Jiang}, \bibinfo{person}{Francis
  Williams}, \bibinfo{person}{Alexey Artemov}, \bibinfo{person}{Evgeny
  Burnaev}, \bibinfo{person}{Marc Alexa}, \bibinfo{person}{Denis Zorin}, {and}
  \bibinfo{person}{Daniele Panozzo}.} \bibinfo{year}{2019}\natexlab{}.
\newblock \showarticletitle{ABC: A Big CAD Model Dataset For Geometric Deep
  Learning}. In \bibinfo{booktitle}{\emph{CVPR}}.
\newblock


\bibitem[\protect\citeauthoryear{Lee, Grey, Ha, Kunz, Jain, Ye, Srinivasa,
  Stilman, and Liu}{Lee et~al\mbox{.}}{2018}]%
        {Lee2018}
\bibfield{author}{\bibinfo{person}{Jeongseok Lee}, \bibinfo{person}{Michael~X
  Grey}, \bibinfo{person}{Sehoon Ha}, \bibinfo{person}{Tobias Kunz},
  \bibinfo{person}{Sumit Jain}, \bibinfo{person}{Yuting Ye},
  \bibinfo{person}{Siddhartha~S Srinivasa}, \bibinfo{person}{Mike Stilman},
  {and} \bibinfo{person}{C~Karen Liu}.} \bibinfo{year}{2018}\natexlab{}.
\newblock \showarticletitle{Dart: Dynamic animation and robotics toolkit}.
\newblock \bibinfo{journal}{\emph{Journal of Open Source Software}}
  \bibinfo{volume}{3}, \bibinfo{number}{22} (\bibinfo{year}{2018}),
  \bibinfo{pages}{500}.
\newblock


\bibitem[\protect\citeauthoryear{Lee, Park, Lee, and Lee}{Lee
  et~al\mbox{.}}{2019}]%
        {Lee2019}
\bibfield{author}{\bibinfo{person}{Seunghwan Lee}, \bibinfo{person}{Moonseok
  Park}, \bibinfo{person}{Kyoungmin Lee}, {and} \bibinfo{person}{Jehee Lee}.}
  \bibinfo{year}{2019}\natexlab{}.
\newblock \showarticletitle{Scalable muscle-actuated human simulation and
  control}.
\newblock \bibinfo{journal}{\emph{ACM Trans. Graph.}} \bibinfo{volume}{38},
  \bibinfo{number}{4} (\bibinfo{year}{2019}).
\newblock


\bibitem[\protect\citeauthoryear{Li, Wu, Tedrake, Tenenbaum, and Torralba}{Li
  et~al\mbox{.}}{2019a}]%
        {Li2019a}
\bibfield{author}{\bibinfo{person}{Yunzhu Li}, \bibinfo{person}{Jiajun Wu},
  \bibinfo{person}{Russ Tedrake}, \bibinfo{person}{Joshua~B. Tenenbaum}, {and}
  \bibinfo{person}{Antonio Torralba}.} \bibinfo{year}{2019}\natexlab{a}.
\newblock \showarticletitle{Learning Particle Dynamics for Manipulating Rigid
  Bodies, Deformable Objects, and Fluids}. In
  \bibinfo{booktitle}{\emph{International Conference on Learning
  Representations}}.
\newblock


\bibitem[\protect\citeauthoryear{Li, Wu, Zhu, Tenenbaum, Torralba, and
  Tedrake}{Li et~al\mbox{.}}{2019b}]%
        {Li2019}
\bibfield{author}{\bibinfo{person}{Yunzhu Li}, \bibinfo{person}{Jiajun Wu},
  \bibinfo{person}{Jun-Yan Zhu}, \bibinfo{person}{Joshua~B Tenenbaum},
  \bibinfo{person}{Antonio Torralba}, {and} \bibinfo{person}{Russ Tedrake}.}
  \bibinfo{year}{2019}\natexlab{b}.
\newblock \showarticletitle{Propagation networks for model-based control under
  partial observation}. In \bibinfo{booktitle}{\emph{2019 International
  Conference on Robotics and Automation}}. IEEE.
\newblock


\bibitem[\protect\citeauthoryear{Liu, Roberson, and Kong}{Liu
  et~al\mbox{.}}{2017}]%
        {Liu2017}
\bibfield{author}{\bibinfo{person}{Chenang Liu}, \bibinfo{person}{David
  Roberson}, {and} \bibinfo{person}{Zhenyu Kong}.}
  \bibinfo{year}{2017}\natexlab{}.
\newblock \showarticletitle{Textural analysis-based online closed-loop quality
  control for additive manufacturing processes}. In
  \bibinfo{booktitle}{\emph{IIE Annual Conference. Proceedings}}. Institute of
  Industrial and Systems Engineers (IISE).
\newblock


\bibitem[\protect\citeauthoryear{Liu and Hodgins}{Liu and Hodgins}{2018}]%
        {Liu2018}
\bibfield{author}{\bibinfo{person}{Libin Liu} {and} \bibinfo{person}{Jessica
  Hodgins}.} \bibinfo{year}{2018}\natexlab{}.
\newblock \showarticletitle{Learning basketball dribbling skills using
  trajectory optimization and deep reinforcement learning}.
\newblock \bibinfo{journal}{\emph{ACM Trans. Graph.}} \bibinfo{volume}{37},
  \bibinfo{number}{4} (\bibinfo{year}{2018}).
\newblock


\bibitem[\protect\citeauthoryear{Ma, Tian, Pan, Ren, and Manocha}{Ma
  et~al\mbox{.}}{2018}]%
        {Ma2018}
\bibfield{author}{\bibinfo{person}{Pingchuan Ma}, \bibinfo{person}{Yunsheng
  Tian}, \bibinfo{person}{Zherong Pan}, \bibinfo{person}{Bo Ren}, {and}
  \bibinfo{person}{Dinesh Manocha}.} \bibinfo{year}{2018}\natexlab{}.
\newblock \showarticletitle{Fluid directed rigid body control using deep
  reinforcement learning}.
\newblock \bibinfo{journal}{\emph{ACM Trans. Graph.}} \bibinfo{volume}{37},
  \bibinfo{number}{4} (\bibinfo{year}{2018}).
\newblock


\bibitem[\protect\citeauthoryear{Macklin and M{\"u}ller}{Macklin and
  M{\"u}ller}{2013}]%
        {Macklin2013}
\bibfield{author}{\bibinfo{person}{Miles Macklin} {and}
  \bibinfo{person}{Matthias M{\"u}ller}.} \bibinfo{year}{2013}\natexlab{}.
\newblock \showarticletitle{Position based fluids}.
\newblock \bibinfo{journal}{\emph{ACM Trans. Graph.}} \bibinfo{volume}{32},
  \bibinfo{number}{4} (\bibinfo{year}{2013}).
\newblock


\bibitem[\protect\citeauthoryear{Marple}{Marple}{1980}]%
        {Marple1980}
\bibfield{author}{\bibinfo{person}{Larry Marple}.}
  \bibinfo{year}{1980}\natexlab{}.
\newblock \showarticletitle{A new autoregressive spectrum analysis algorithm}.
\newblock \bibinfo{journal}{\emph{IEEE Transactions on Acoustics, Speech, and
  Signal Processing}} \bibinfo{volume}{28}, \bibinfo{number}{4}
  (\bibinfo{year}{1980}), \bibinfo{pages}{441--454}.
\newblock


\bibitem[\protect\citeauthoryear{Mnih, Kavukcuoglu, Silver, Rusu, Veness,
  Bellemare, Graves, Riedmiller, Fidjeland, Ostrovski, et~al\mbox{.}}{Mnih
  et~al\mbox{.}}{2015}]%
        {Mnih2015}
\bibfield{author}{\bibinfo{person}{Volodymyr Mnih}, \bibinfo{person}{Koray
  Kavukcuoglu}, \bibinfo{person}{David Silver}, \bibinfo{person}{Andrei~A
  Rusu}, \bibinfo{person}{Joel Veness}, \bibinfo{person}{Marc~G Bellemare},
  \bibinfo{person}{Alex Graves}, \bibinfo{person}{Martin Riedmiller},
  \bibinfo{person}{Andreas~K Fidjeland}, \bibinfo{person}{Georg Ostrovski},
  {et~al\mbox{.}}} \bibinfo{year}{2015}\natexlab{}.
\newblock \showarticletitle{Human-level control through deep reinforcement
  learning}.
\newblock \bibinfo{journal}{\emph{Nature}} \bibinfo{volume}{518},
  \bibinfo{number}{7540} (\bibinfo{year}{2015}), \bibinfo{pages}{529--533}.
\newblock


\bibitem[\protect\citeauthoryear{Mozaffar, Paul, Al-Bahrani, Wolff, Choudhary,
  Agrawal, Ehmann, and Cao}{Mozaffar et~al\mbox{.}}{2018}]%
        {Mozaffar2018}
\bibfield{author}{\bibinfo{person}{Mojtaba Mozaffar}, \bibinfo{person}{Arindam
  Paul}, \bibinfo{person}{Reda Al-Bahrani}, \bibinfo{person}{Sarah Wolff},
  \bibinfo{person}{Alok Choudhary}, \bibinfo{person}{Ankit Agrawal},
  \bibinfo{person}{Kornel Ehmann}, {and} \bibinfo{person}{Jian Cao}.}
  \bibinfo{year}{2018}\natexlab{}.
\newblock \showarticletitle{Data-driven prediction of the high-dimensional
  thermal history in directed energy deposition processes via recurrent neural
  networks}.
\newblock \bibinfo{journal}{\emph{Manufacturing letters}}  \bibinfo{volume}{18}
  (\bibinfo{year}{2018}), \bibinfo{pages}{35--39}.
\newblock


\bibitem[\protect\citeauthoryear{M{\"u}ller, Charypar, and Gross}{M{\"u}ller
  et~al\mbox{.}}{2003}]%
        {Muller2003}
\bibfield{author}{\bibinfo{person}{Matthias M{\"u}ller}, \bibinfo{person}{David
  Charypar}, {and} \bibinfo{person}{Markus~H Gross}.}
  \bibinfo{year}{2003}\natexlab{}.
\newblock \showarticletitle{Particle-based fluid simulation for interactive
  applications.}. In \bibinfo{booktitle}{\emph{Symposium on Computer
  animation}}.
\newblock


\bibitem[\protect\citeauthoryear{M{\"u}ller, Heidelberger, Hennix, and
  Ratcliff}{M{\"u}ller et~al\mbox{.}}{2007}]%
        {Muller2007}
\bibfield{author}{\bibinfo{person}{Matthias M{\"u}ller}, \bibinfo{person}{Bruno
  Heidelberger}, \bibinfo{person}{Marcus Hennix}, {and} \bibinfo{person}{John
  Ratcliff}.} \bibinfo{year}{2007}\natexlab{}.
\newblock \showarticletitle{Position based dynamics}.
\newblock \bibinfo{journal}{\emph{Journal of Visual Communication and Image
  Representation}} \bibinfo{volume}{18}, \bibinfo{number}{2}
  (\bibinfo{year}{2007}).
\newblock


\bibitem[\protect\citeauthoryear{Nagabandi, Kahn, Fearing, and
  Levine}{Nagabandi et~al\mbox{.}}{2018}]%
        {Nagabandi2018}
\bibfield{author}{\bibinfo{person}{Anusha Nagabandi}, \bibinfo{person}{Gregory
  Kahn}, \bibinfo{person}{Ronald~S Fearing}, {and} \bibinfo{person}{Sergey
  Levine}.} \bibinfo{year}{2018}\natexlab{}.
\newblock \showarticletitle{Neural network dynamics for model-based deep
  reinforcement learning with model-free fine-tuning}. In
  \bibinfo{booktitle}{\emph{ICRA 2018}}. IEEE.
\newblock


\bibitem[\protect\citeauthoryear{Ogoke and Farimani}{Ogoke and
  Farimani}{2021}]%
        {Ogoke2021}
\bibfield{author}{\bibinfo{person}{Francis Ogoke} {and}
  \bibinfo{person}{Amir~Barati Farimani}.} \bibinfo{year}{2021}\natexlab{}.
\newblock \showarticletitle{Thermal control of laser powder bed fusion using
  deep reinforcement learning}.
\newblock \bibinfo{journal}{\emph{Additive Manufacturing}}
  \bibinfo{volume}{46} (\bibinfo{year}{2021}).
\newblock
\showISSN{2214-8604}


\bibitem[\protect\citeauthoryear{Oh, Singh, and Lee}{Oh et~al\mbox{.}}{2017}]%
        {Oh2017}
\bibfield{author}{\bibinfo{person}{Junhyuk Oh}, \bibinfo{person}{Satinder
  Singh}, {and} \bibinfo{person}{Honglak Lee}.}
  \bibinfo{year}{2017}\natexlab{}.
\newblock \showarticletitle{Value Prediction Network}. In
  \bibinfo{booktitle}{\emph{NIPS}}.
\newblock


\bibitem[\protect\citeauthoryear{Peng, Abbeel, Levine, and van~de Panne}{Peng
  et~al\mbox{.}}{2018}]%
        {Peng2018}
\bibfield{author}{\bibinfo{person}{Xue~Bin Peng}, \bibinfo{person}{Pieter
  Abbeel}, \bibinfo{person}{Sergey Levine}, {and} \bibinfo{person}{Michiel
  van~de Panne}.} \bibinfo{year}{2018}\natexlab{}.
\newblock \showarticletitle{Deepmimic: Example-guided deep reinforcement
  learning of physics-based character skills}.
\newblock \bibinfo{journal}{\emph{ACM Trans. Graph.}} \bibinfo{volume}{37},
  \bibinfo{number}{4} (\bibinfo{year}{2018}).
\newblock


\bibitem[\protect\citeauthoryear{Rajeswaran, Kumar, Gupta, Vezzani, Schulman,
  Todorov, and Levine}{Rajeswaran et~al\mbox{.}}{2017}]%
        {Rajeswaran2017}
\bibfield{author}{\bibinfo{person}{Aravind Rajeswaran}, \bibinfo{person}{Vikash
  Kumar}, \bibinfo{person}{Abhishek Gupta}, \bibinfo{person}{Giulia Vezzani},
  \bibinfo{person}{John Schulman}, \bibinfo{person}{Emanuel Todorov}, {and}
  \bibinfo{person}{Sergey Levine}.} \bibinfo{year}{2017}\natexlab{}.
\newblock \showarticletitle{Learning complex dexterous manipulation with deep
  reinforcement learning and demonstrations}.
\newblock \bibinfo{journal}{\emph{arXiv preprint arXiv:1709.10087}}
  (\bibinfo{year}{2017}).
\newblock


\bibitem[\protect\citeauthoryear{Schenck and Fox}{Schenck and Fox}{2018}]%
        {Schenck2018}
\bibfield{author}{\bibinfo{person}{Connor Schenck} {and}
  \bibinfo{person}{Dieter Fox}.} \bibinfo{year}{2018}\natexlab{}.
\newblock \showarticletitle{Spnets: Differentiable fluid dynamics for deep
  neural networks}. In \bibinfo{booktitle}{\emph{Conference on Robot
  Learning}}. PMLR, \bibinfo{pages}{317--335}.
\newblock


\bibitem[\protect\citeauthoryear{Schulman, Wolski, Dhariwal, Radford, and
  Klimov}{Schulman et~al\mbox{.}}{2017}]%
        {Schulman2017}
\bibfield{author}{\bibinfo{person}{John Schulman}, \bibinfo{person}{Filip
  Wolski}, \bibinfo{person}{Prafulla Dhariwal}, \bibinfo{person}{Alec Radford},
  {and} \bibinfo{person}{Oleg Klimov}.} \bibinfo{year}{2017}\natexlab{}.
\newblock \showarticletitle{Proximal policy optimization algorithms}.
\newblock \bibinfo{journal}{\emph{arXiv preprint arXiv:1707.06347}}
  (\bibinfo{year}{2017}).
\newblock


\bibitem[\protect\citeauthoryear{Silver, Hasselt, Hessel, Schaul, Guez, Harley,
  Dulac-Arnold, Reichert, Rabinowitz, Barreto, et~al\mbox{.}}{Silver
  et~al\mbox{.}}{2017}]%
        {Silver2017}
\bibfield{author}{\bibinfo{person}{David Silver}, \bibinfo{person}{Hado
  Hasselt}, \bibinfo{person}{Matteo Hessel}, \bibinfo{person}{Tom Schaul},
  \bibinfo{person}{Arthur Guez}, \bibinfo{person}{Tim Harley},
  \bibinfo{person}{Gabriel Dulac-Arnold}, \bibinfo{person}{David Reichert},
  \bibinfo{person}{Neil Rabinowitz}, \bibinfo{person}{Andre Barreto},
  {et~al\mbox{.}}} \bibinfo{year}{2017}\natexlab{}.
\newblock \showarticletitle{The predictron: End-to-end learning and planning}.
  In \bibinfo{booktitle}{\emph{International Conference on Machine Learning}}.
  PMLR, \bibinfo{pages}{3191--3199}.
\newblock


\bibitem[\protect\citeauthoryear{Sitthi-Amorn, Ramos, Wangy, Kwan, Lan, Wang,
  and Matusik}{Sitthi-Amorn et~al\mbox{.}}{2015}]%
        {Sitthi2015}
\bibfield{author}{\bibinfo{person}{Pitchaya Sitthi-Amorn},
  \bibinfo{person}{Javier~E Ramos}, \bibinfo{person}{Yuwang Wangy},
  \bibinfo{person}{Joyce Kwan}, \bibinfo{person}{Justin Lan},
  \bibinfo{person}{Wenshou Wang}, {and} \bibinfo{person}{Wojciech Matusik}.}
  \bibinfo{year}{2015}\natexlab{}.
\newblock \showarticletitle{MultiFab: a machine vision assisted platform for
  multi-material 3D printing}.
\newblock \bibinfo{journal}{\emph{ACM Trans. Graph.}} \bibinfo{volume}{34},
  \bibinfo{number}{4} (\bibinfo{year}{2015}), \bibinfo{pages}{1--11}.
\newblock


\bibitem[\protect\citeauthoryear{Srinivas, Jabri, Abbeel, Levine, and
  Finn}{Srinivas et~al\mbox{.}}{2018}]%
        {Srinivas2018}
\bibfield{author}{\bibinfo{person}{Aravind Srinivas}, \bibinfo{person}{Allan
  Jabri}, \bibinfo{person}{Pieter Abbeel}, \bibinfo{person}{Sergey Levine},
  {and} \bibinfo{person}{Chelsea Finn}.} \bibinfo{year}{2018}\natexlab{}.
\newblock \showarticletitle{Universal planning networks: Learning generalizable
  representations for visuomotor control}. In
  \bibinfo{booktitle}{\emph{International Conference on Machine Learning}}.
  PMLR, \bibinfo{pages}{4732--4741}.
\newblock


\bibitem[\protect\citeauthoryear{Tang, Tan, and Wong}{Tang
  et~al\mbox{.}}{2018}]%
        {Tang2018}
\bibfield{author}{\bibinfo{person}{Chao Tang}, \bibinfo{person}{Jie~Lun Tan},
  {and} \bibinfo{person}{Chee~How Wong}.} \bibinfo{year}{2018}\natexlab{}.
\newblock \showarticletitle{A numerical investigation on the physical
  mechanisms of single track defects in selective laser melting}.
\newblock \bibinfo{journal}{\emph{International Journal of Heat and Mass
  Transfer}}  \bibinfo{volume}{126} (\bibinfo{year}{2018}),
  \bibinfo{pages}{957--968}.
\newblock


\bibitem[\protect\citeauthoryear{Todorov, Erez, and Tassa}{Todorov
  et~al\mbox{.}}{2012}]%
        {Todorov2012}
\bibfield{author}{\bibinfo{person}{Emanuel Todorov}, \bibinfo{person}{Tom
  Erez}, {and} \bibinfo{person}{Yuval Tassa}.} \bibinfo{year}{2012}\natexlab{}.
\newblock \showarticletitle{Mujoco: A physics engine for model-based control}.
  In \bibinfo{booktitle}{\emph{2012 IEEE/RSJ International Conference on
  Intelligent Robots and Systems}}. IEEE, \bibinfo{pages}{5026--5033}.
\newblock


\bibitem[\protect\citeauthoryear{Toussaint, Allen, Smith, and
  Tenenbaum}{Toussaint et~al\mbox{.}}{2018}]%
        {Toussaint2018}
\bibfield{author}{\bibinfo{person}{Marc~A Toussaint},
  \bibinfo{person}{Kelsey~Rebecca Allen}, \bibinfo{person}{Kevin~A Smith},
  {and} \bibinfo{person}{Joshua~B Tenenbaum}.} \bibinfo{year}{2018}\natexlab{}.
\newblock \showarticletitle{Differentiable physics and stable modes for
  tool-use and manipulation planning}.
\newblock \bibinfo{journal}{\emph{Robotics: Science and Systems Foundation}}
  (\bibinfo{year}{2018}).
\newblock


\bibitem[\protect\citeauthoryear{Wang, Tan, Liu, and Tor}{Wang
  et~al\mbox{.}}{2018}]%
        {Wang2018}
\bibfield{author}{\bibinfo{person}{Chengcheng Wang}, \bibinfo{person}{Xipeng
  Tan}, \bibinfo{person}{Erjia Liu}, {and} \bibinfo{person}{Shu~Beng Tor}.}
  \bibinfo{year}{2018}\natexlab{}.
\newblock \showarticletitle{Process parameter optimization and mechanical
  properties for additively manufactured stainless steel 316L parts by
  selective electron beam melting}.
\newblock \bibinfo{journal}{\emph{Materials \& Design}}  \bibinfo{volume}{147}
  (\bibinfo{year}{2018}).
\newblock


\bibitem[\protect\citeauthoryear{Wang, Tan, Tor, and Lim}{Wang
  et~al\mbox{.}}{2020}]%
        {Wang2020}
\bibfield{author}{\bibinfo{person}{Chengcheng Wang}, \bibinfo{person}{XP Tan},
  \bibinfo{person}{SB Tor}, {and} \bibinfo{person}{CS Lim}.}
  \bibinfo{year}{2020}\natexlab{}.
\newblock \showarticletitle{Machine learning in additive manufacturing:
  State-of-the-art and perspectives}.
\newblock \bibinfo{journal}{\emph{Additive Manufacturing}}
  (\bibinfo{year}{2020}).
\newblock


\bibitem[\protect\citeauthoryear{Wu, Yan, Kurutach, Pinto, and Abbeel}{Wu
  et~al\mbox{.}}{2019}]%
        {Wu2019}
\bibfield{author}{\bibinfo{person}{Yilin Wu}, \bibinfo{person}{Wilson Yan},
  \bibinfo{person}{Thanard Kurutach}, \bibinfo{person}{Lerrel Pinto}, {and}
  \bibinfo{person}{Pieter Abbeel}.} \bibinfo{year}{2019}\natexlab{}.
\newblock \showarticletitle{Learning to manipulate deformable objects without
  demonstrations}.
\newblock \bibinfo{journal}{\emph{arXiv preprint arXiv:1910.13439}}
  (\bibinfo{year}{2019}).
\newblock


\bibitem[\protect\citeauthoryear{Xu, Du, Foshey, Li, Zhu, Schulz, and
  Matusik}{Xu et~al\mbox{.}}{2019}]%
        {xu2019learning}
\bibfield{author}{\bibinfo{person}{Jie Xu}, \bibinfo{person}{Tao Du},
  \bibinfo{person}{Michael Foshey}, \bibinfo{person}{Beichen Li},
  \bibinfo{person}{Bo Zhu}, \bibinfo{person}{Adriana Schulz}, {and}
  \bibinfo{person}{Wojciech Matusik}.} \bibinfo{year}{2019}\natexlab{}.
\newblock \showarticletitle{Learning to fly: computational controller design
  for hybrid UAVs with reinforcement learning}.
\newblock \bibinfo{journal}{\emph{ACM Trans. Graph.}} \bibinfo{volume}{38},
  \bibinfo{number}{4} (\bibinfo{year}{2019}).
\newblock


\bibitem[\protect\citeauthoryear{Yan, Qian, Ge, Lin, Liu, Lin, and Wagner}{Yan
  et~al\mbox{.}}{2018}]%
        {Yan2018}
\bibfield{author}{\bibinfo{person}{Wentao Yan}, \bibinfo{person}{Ya Qian},
  \bibinfo{person}{Wenjun Ge}, \bibinfo{person}{Stephen Lin},
  \bibinfo{person}{Wing~Kam Liu}, \bibinfo{person}{Feng Lin}, {and}
  \bibinfo{person}{Gregory~J Wagner}.} \bibinfo{year}{2018}\natexlab{}.
\newblock \showarticletitle{Meso-scale modeling of multiple-layer fabrication
  process in selective electron beam melting: inter-layer/track voids
  formation}.
\newblock \bibinfo{journal}{\emph{Materials \& Design}}  \bibinfo{volume}{141}
  (\bibinfo{year}{2018}), \bibinfo{pages}{210--219}.
\newblock


\bibitem[\protect\citeauthoryear{Yao, Imani, and Yang}{Yao
  et~al\mbox{.}}{2018}]%
        {Yao2018}
\bibfield{author}{\bibinfo{person}{Bing Yao}, \bibinfo{person}{Farhad Imani},
  {and} \bibinfo{person}{Hui Yang}.} \bibinfo{year}{2018}\natexlab{}.
\newblock \showarticletitle{Markov decision process for image-guided additive
  manufacturing}.
\newblock \bibinfo{journal}{\emph{IEEE Robotics and Automation Letters}}
  \bibinfo{volume}{3}, \bibinfo{number}{4} (\bibinfo{year}{2018}),
  \bibinfo{pages}{2792--2798}.
\newblock


\bibitem[\protect\citeauthoryear{Yu, Park, and Lee}{Yu et~al\mbox{.}}{2019}]%
        {Yu2019}
\bibfield{author}{\bibinfo{person}{Ri Yu}, \bibinfo{person}{Hwangpil Park},
  {and} \bibinfo{person}{Jehee Lee}.} \bibinfo{year}{2019}\natexlab{}.
\newblock \showarticletitle{Figure skating simulation from video}. In
  \bibinfo{booktitle}{\emph{Computer graphics forum}},
  Vol.~\bibinfo{volume}{38}. Wiley Online Library, \bibinfo{pages}{225--234}.
\newblock


\bibitem[\protect\citeauthoryear{Zhang, Yu, Liu, Kemp, and Turk}{Zhang
  et~al\mbox{.}}{2020}]%
        {Zhang2020}
\bibfield{author}{\bibinfo{person}{Yunbo Zhang}, \bibinfo{person}{Wenhao Yu},
  \bibinfo{person}{C~Karen Liu}, \bibinfo{person}{Charlie Kemp}, {and}
  \bibinfo{person}{Greg Turk}.} \bibinfo{year}{2020}\natexlab{}.
\newblock \showarticletitle{Learning to manipulate amorphous materials}.
\newblock \bibinfo{journal}{\emph{ACM Trans. Graph.}} \bibinfo{volume}{39},
  \bibinfo{number}{6} (\bibinfo{year}{2020}).
\newblock


\bibitem[\protect\citeauthoryear{Zhou and Jacobson}{Zhou and Jacobson}{2016}]%
        {Thingi10K}
\bibfield{author}{\bibinfo{person}{Qingnan Zhou} {and} \bibinfo{person}{Alec
  Jacobson}.} \bibinfo{year}{2016}\natexlab{}.
\newblock \showarticletitle{Thingi10K: A Dataset of 10,000 3D-Printing Models}.
\newblock \bibinfo{journal}{\emph{arXiv preprint arXiv:1605.04797}}
  (\bibinfo{year}{2016}).
\newblock


\end{thebibliography}

\appendix

\section{Introduction}

The appendix describes additional details about the printing hardware (Section~\ref{sec:hardware}), numerical simulation (Section~\ref{sec:simulation}), and reinforcement learning formulation (Section~\ref{sec:learning}). We also include the full physical results manufactured for this paper, (Section~\ref{sec:results}).

\section{Hardware Apparatus}\label{sec:hardware}

We developed a direct write 3D printing platform with an optical feedback system that can measure the dispensed material real-time, in-situ. The 3D printer is comprised of a pressure-driven syringe pump and pressure controller (Enfield Technologies), a 3-axis Cartesian robot (Hiwin KK), an optical imaging system, a back-lit build platform, 3D-printer controller, and CPU, (Figure~\ref{fig:hardware_full}). The 3-axis Cartesian robot is used to locate the build platform in the x and y-direction and the print carriage in the z-direction. The pressure-driven syringe pump and pressure controller are used to dispense an optically translucent material onto the back-lit build platform. The back-lit platform is used to illuminate the dispensed material. The movement of the robot, actuation of the syringe pump, and timing of the cameras are controlled via the controller. The CPU is used to process the images after they are acquired and compute updated commands to send to the controller.

\myfigure{hardware_full}{
    The printing apparatus consisting of a 3-axis Cartesian robot, a direct write printing head, a camera setup, illuminated build plate, calibration facilities, and a orange UV-cover.
}

\subsection{Vision Module}

To enable realtime control of the printing process, we implemented an in-situ view of the material deposition. Due to the occlusions caused by the dispensing nozzle, no single camera can capture the full view. Therefore, we opted for a two-camera setup. More specifically, we place two CMOS cameras (Basler AG, Ahrensburg, Germany) at 45 degrees on each side of the dispensing nozzle, (Figure~\ref{fig:hardware_full}). We further process the images from the cameras to obtain a single top-down view of the deposition. We start by calibrating the camera by collecting a set of images and estimating its intrinsic parameters, (Figure~\ref{fig:imaging_setup} calibration). To obtain a single top-down view, we capture a calibration target aligned with the image frames of both cameras, (Figure~\ref{fig:imaging_setup} homography). We can stitch the images into a single view from an over-the-top virtual camera by calculating the homography between the captured targets and an ideal top-down view. Finally, we mask the location of each nozzle in the image (Figure~\ref{fig:imaging_setup} nozzle masks) and obtain the final in-situ view, (Figure~\ref{fig:imaging_setup} stitched image).

\myfigure{imaging_setup}{
    The calibration of the imaging setup. First intrinsic parameters are estimated from calibration patterns. Next, we compute the extrinsic calibration by calculating homographies between the cameras and an overhead view. We extract the masks by thresholding a photo of the nozzle. The final stitched image consists of 4 regions: (1) view only in the left camera, (2) view only in the right camera, (3) view in both cameras, (4) view in no camera. The final stitched image is shown on the right.
}

To extract the thickness of the deposited material, we rely on its translucency properties. More precisely, we correlate the material thickness with its optical intensity, (Figure~\ref{fig:thickness_calibration}). We do this by depositing the material at various thicknesses and taking a picture with our camera setup. The optical intensity then decays exponentially with an increased thickness.

\myfigure{thickness_calibration}{
    Calibration images for correlating deposited material thickness with optical intensity and the corresponding fit.
}

\subsection{Mechanical Calibration}

In our simulation, we assume the needle is centered with respect to the in-situ view. To ensure that this assumption holds with the physical hardware, we calibrate the location of the dispensing needle within the field of view of each camera and with respect to the build platform. First, a dial indicator is used to measure the height of the nozzle in z, and the fine adjustment stage is adjusted until the nozzle is 254 microns above the print platform. Next, using a calibration target located on the build platform and the fine adjustment stage, the nozzle is centered in the field of view of each camera. This calibration procedure is done each time the nozzle is replaced during the start of each printing session. 

\section{Simulation Details}\label{sec:simulation}

The discretization choice of the numerical model affects the learning process. We experimented with two options: (1) time-based and (2) distance-based. We originally experimented with time-based discretization. However, we found out that time discretization is not suitable for printer modeling. As the velocity in simulation approaches zero, the difference in deposited material becomes progressively smaller until the gradient information completely vanishes, (Figure~\ref{fig:discretization_choice_inset} top). Moreover, a time-based discretization allows the policy to affect the number of evaluations of the environment directly. As a result, it can avoid being punished for bad material deposition by quickly rushing the environment to finish. Considering these factors we opted for distance-based discretization, (Figure~\ref{fig:discretization_choice_inset} bottom). The policy specifies the desired velocity at each interaction point, and the environment travels a predefined distance (0.315 mm) at the desired speed. This helps to regularize the reward function and enables learning of varying control policies.

\mySfigure{discretization_choice_inset}{0.5}{
    The amount of material deposited between two simulation steps depends on the discretization choice. With time-based discretization, the reward gradient vanishes at low speeds. In contrast, distance-based discretization produces a more uniform response.
}

An interesting design element is the orientation of the control polygons created by the slicer. When the outline is defined as points given counter-clockwise, then due to the applied rotation, each view is split roughly into two half-spaces, (Figure~\ref{fig:path_orientation_inset}). The bottom one corresponds to outside \ie generally black, and the upper one corresponds to inside \ie generally white. However, the situation changes when outlining a hole. When printing a hole the two half-spaces swap location. We can remove this disambiguity by changing the orientation of the polylines defining holes in the model. By orienting them clockwise, we will effectively swap the two half-spaces to the same orientation as when printing the outer part. As a result, we achieve a better usage of trajectories and a more robust control scheme that does not need to be separately trained for each print's outer and inner parts.

\mySfigure{path_orientation_inset}{0.5}{
    The observation space is split into two half-spaces corresponding to the target printout and void. Therefore, it is advantageous to pick different printing directions for objects' outlines and holes inside to minimize the observation space and reuse trajectories. By picking different directions, we guarantee that material is always deposited only on one side.
}

\section{Reinforcement Learning Framework}\label{sec:learning}

To train our control policy, we start with a g-code generated by a slicer. As inputs to the slicer, we consider a set of 3D models collected from the Thingy10k dataset. To train a controller, the input models need to be carefully selected. On the one hand, if we pick an object with too low-frequency features with respect to the printing nozzle size, then any printing errors due to control policy will have negligible influence on the final result. On the other hand, if we pick a model with too high-frequency features regarding the printing nozzle, the nozzle will be physically unable to reproduce these features. As a result, we opted for a manual selection of 18 models that span a wide variety of features, (Figure~\ref{fig:curriculum}). Each model is scaled to fit into a printing volume of $22\times 22$ mm and sliced at random locations.

\myfigure{curriculum}{
    Models contained in our training curriculum.
}

Our policy is represented as a CNN modeled after \citet{Mnih2015}. The network input is a $84\times 84 \times 3$ image. The image is passed through three hidden layers. The convolution layers have the respective parameters: (32 filters, filter size 8, stride 4), (64 filters, filter size 4, stride 2), and (64 filters, filter size 3, stride 1). The final convolved image is linearized and passed through a fully-connected layer with 512 neurons connected to the output actions. Each hidden layer uses the nonlinear rectifier activation. We formulate our objective function as in~\citep{Schulman2017}:
\begin{equation}
    \argmax_\theta \mathbb{E}_t^\mathcal{C}\left[\frac{\pi_{\theta_t}(a_t | s_t)}{\pi_{\theta_{t-1}}(a_t | s_t)}\hat{A}_t\right] ,
    \label{eq:our_objective}
\end{equation}
where $t$ is a timestep in the optimization, $\theta$ are the hyperparameters of a neural network encoding our policy $\pi$ that generates an action $a_t$ based on a set of observations $s_t$, $\hat{A}_t$ is the estimator of the advantage function and the expectation $\mathbb{E}_t^\mathcal{C}$ is an average of a finite batch of samples generated by printing sliced models from our curriculum $\mathcal{C}$. To maximize Equation~\ref{eq:our_objective} we use Principal Policy Optimization (PPO) algorithm~\citep{Schulman2017}.

\subsection{Learning Curves}

We conducted several ablation studies during which we observed the convergence of our learning process. When experimenting with the observation space, we did not observe a significant difference in learning convergence between our full observation space and its reduced variants, (Figure~\ref{fig:training_curves}). During our experiments with action space, we found that a policy that optimizes only the velocity of the nozzle can learn faster than a policy that adjusts the deposition path, (Figure~\ref{fig:training_curves_ablation} left). Lastly, we observed the convergence rate between different reward computation strategies, (Figure~\ref{fig:training_curves_ablation} right). The delayed reward converges significantly slower than a reward function with instantaneous feedback, and it is unclear if a performance similar to our privileged reward can be achieved.

\myfigure{training_curves}{
    Training curves for controllers with constant material flow.
}

\myfigure{training_curves_ablation}{
    Training curves for controllers with variable material flow.
}

\section{Results}\label{sec:results}

For evaluation, we constructed a dataset consisting of freeform and CAD geometries that were not present in training. A subset of the dataset is visualized in Figure~\ref{fig:evaluation}.

\myfigure{evaluation}{
    Exemplar models from the evaluation dataset.
}

\subsection{Slices Used to Estimate Outline Improvement}

We estimate the quality of deposition by evaluating the under and over deposition histograms on a subset of the evaluation dataset, (Figure~\ref{fig:sim_histograms_all}).

\myfigure{sim_histograms_all}{
    Recovered histograms for a subset of slices from the evaluation dataset show tighter deposition of material achieved by our closed-loop control policy.
}

\newpage

\subsection{Detailed Physical Results}

We fabricated 11 shapes from the training and 11 shapes from the evaluation dataset on a physical apparatus using a low and high viscosity material, (Figure~\ref{fig:physical_results_digital}).

\mycSfigure{physical_results_digital}{0.93}{
    Policy evaluation on physical hardware.
}

\end{document}